\documentclass[aps,prl,twocolumn,groupedaddress, showpacs]{revtex4-1}
\usepackage{graphicx}
\usepackage[ansinew]{inputenc}
\usepackage{colortbl}

\newcommand{\e}{\varepsilon}

\begin{document}

\title{Dynamics of Protein Hydration Water}


\author{M. Wolf}
\email[Corresponding{\color{white}a}author: ]{Martin.Wolf@physik.uni-augsburg.de}
\author{S. Emmert}
\author{R. Gulich}
\author{P. Lunkenheimer}
\author{A. Loidl}
\affiliation{Experimental Physics V, Center for Electronic Correlations and Magnetism, University of Augsburg, Universitätsstr. 2, 86135 Augsburg, Germany}



\begin{abstract}
We present the frequency- and temperature-dependent dielectric properties of lysozyme solutions in a broad concentration regime, measured at subzero temperatures and compare the results with measurements above the freezing point of water and on hydrated lysozyme powder. Our experiments allow examining the dynamics of unfreezable hydration water in a broad temperature range including the so-called No Man's Land (160--235\,K). The obtained results prove the bimodality of the hydration shell dynamics and are discussed in the context of the highly-debated fragile-to-strong transition of water.
\end{abstract}


\maketitle
Water is essential for nearly all biologically active systems. Prominent examples are globular proteins, whose functional and physical properties are fundamentally determined by the presence of water \cite{Gregory1994, Kuntz1974}. Especially, the so-called ``hydration water'', i.e. the shell of water molecules in the close vicinity of the protein surface, strongly interacts with the latter. As a consequence, hydration water does not crystallize, even at temperatures far below water's nominal freezing point. Due to their importance, protein-solvent interactions are a very active field of research \cite{Gregory1994, Doster2010a, Cametti2013, Ngai2011, Ngai2013, Nibali2014, McMahon2014}. It is commonly believed that the water molecules of the hydration shell cause a relaxation process similar to that of pure water but slowed down due to the bonding to the protein. However, it is a matter of debate if there are \emph{two} such hydration shell relaxations \cite{Grant1986, Bone1992, Knocks2001, Oleinikova2004, Cametti2011, Wolf2012}, which would be in accordance with the idea that there are at least two layers of hydration water with different bonding energies. This is difficult to decide because proteins exhibit numerous intramolecular motions and solvent interactions, all giving rise to relaxation processes complicating the dielectric spectra of protein solutions.

\begin{figure}[h]
\includegraphics[width=0.75\linewidth]{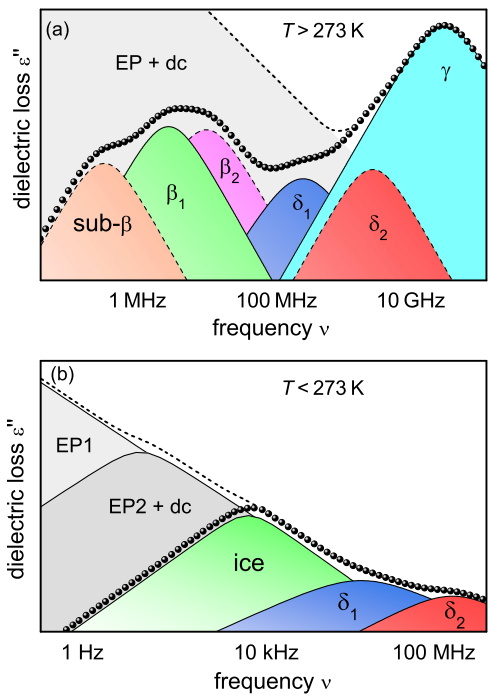}%
\caption{\label{fig0}Schematic plot of the dielectric loss of protein solutions above (a) and below (b) the freezing point of water. The circles show the loss after subtraction of the dc conductivity and EP contributions (grey regions). Frame (a) comprises the major processes hitherto found in protein solutions \cite{WolfDiss}: The commonly observed $\beta$- and $\gamma$-relaxations are due to the reorientational motions of the dipolar protein and water molecules, respectively. The $\delta$-relaxations are ascribed to hydration water. Additional contributions may arise from conformational sampling (sub-$\beta$). Frozen lysozyme solutions (b) show three intrinsic relaxations: The ``ice'' process results from proton hopping in the ice matrix. In the frequency range of the $\delta$-dispersion, two relaxations are depicted ($\delta_1$, $\delta_2$).}
\end{figure}

Figure \ref{fig0}(a) schematically shows the contributions of the major relaxation processes of protein solutions to dielectric loss spectra, $\varepsilon''(\nu)$. Here one should be aware that most of them only show up after subtraction of strong ionic conductivity and electrode-polarization (EP) effects (grey region), typical for ion-conducting materials \cite{Macdonald1987a, Feldman2003, Pimenov1998, Gulich2009, Emmert2011, Wolf2011}.  Protein solutions always show one or two $\beta$-relaxations (using the biophysical nomenclature), resulting from the tumbling of the dipolar protein, and a very strong $\gamma$-relaxation caused by the dipolar water molecules \cite{Perrin1934, Koenig1975, Essex1977, FEDOTOV1990, Ermolina1998, Knocks2001, Oleinikova2004, Wolf2012}. In addition, they may exhibit proton-fluctuation processes (in the frequency range of the $\beta$-relaxation) \cite{South1972} or sub-$\beta$-relaxations, caused by conformational sampling events \cite{Ban2011}.
The hydration-shell relaxations are expected to be located in the so-called $\delta$-dispersion region between the $\beta$- and  $\gamma$-relaxations. As pointed out earlier \cite{Wolf2012}, a single relaxation function sufficiently describes the $\delta$-dispersion region of protein solutions and, according to Occam's razor, there is no reason to employ a further relaxation function to account for the experimental data. On the other hand, a proper analysis of the $\delta$-dispersion is difficult as large parts of the relevant frequency region are dominated by the $\beta$- and $\gamma$-relaxations [Fig. \ref{fig0}(a)]. In addition, other contributions as protein side-chain motions or internal protein motions may also contribute in this frequency range \cite{Bone1987, Grant1978, Essex1977, Oleinikova2004, PENNOCK1969}.
An often used alternative approach is the investigation of hydrated protein powders \cite{Doster1989, Chen2006, Khodadadi2008, Doster2010a}. However, their dynamics also reveals numerous relaxations, with some of them being extremely hydration-dependent \cite{Tombari2013, Khodadadi2008a, Pizzitutti2001, Pethig1992, Rupley1991}. Thus, the interpretation of hydrated protein-powder studies is difficult, too.

The approach of the present work is to circumvent these problems by investigating protein solutions below the freezing point of water. In this way, the interfering $\beta$- and $\gamma$-relaxations are eliminated as the protein and water molecules cannot easily reorient in the frozen sample and also the conductivity is strongly reduced. In contrast, as the hydration-shell water remains amorphous, the $\delta$-relaxation(s) should still be observable.
The typical $\varepsilon''$ spectrum of a frozen protein solution is shown in Fig. \ref{fig0}(b). The strong relaxation-like processes at low frequencies (EP1, EP2) arise from dc conductivity and non-intrinsic EP effects. The relaxation termed ``ice'' is caused by Bjerrum and ionic defects in the ice structure resulting in proton-hopping processes, physically equivalent to reorientations of water molecules \cite{Johari1981b, Petrenko2006}. In the $\delta$-dispersion region, two $\delta$-relaxation processes are indicated ($\delta_1$, $\delta_2$).

In the present study, we investigate various frozen lysozyme solutions and compare the results with those of protein solutions above the freezing point \cite{Wolf2012} and with a hydrated lysozyme powder.
This gives valuable insight into the hydration-shell dynamics of proteins in a broad temperature range and helps clarifying the question if there are indeed two $\delta$-relaxations. Moreover, the data are discussed in the context of the highly debated No Man's Land and the controversial fragile-to-strong transition of bulk water \cite{Ito1999, Johari2000, Debenedetti2003, Chen2006, Angell2008, Khodadadi2008, vogel2008, Doster2010a, Nakanishi2012}.

The complex dielectric permittivity and conductivity
were determined using two measurement devices
covering the frequency range between $\approx$\,0.1\,Hz and 3\,GHz \cite{supp,Boehmer1989,Schneider2001}.
Lysozyme/water solutions between 3\,mmol/l and 100\,mmol/l were prepared by dissolving
weighed amounts of commercially available lysozyme powder in deionized H$_2$O.
The hydrated powder was prepared by exposing lysozyme powder
to an atmosphere of defined relative humidity.
The degree of hydration was determined to be $h=30$\,wt\%. More experimental details are provided in the Supplemental Material \cite{supp}.

\begin{figure}
\includegraphics[width=0.85\linewidth]{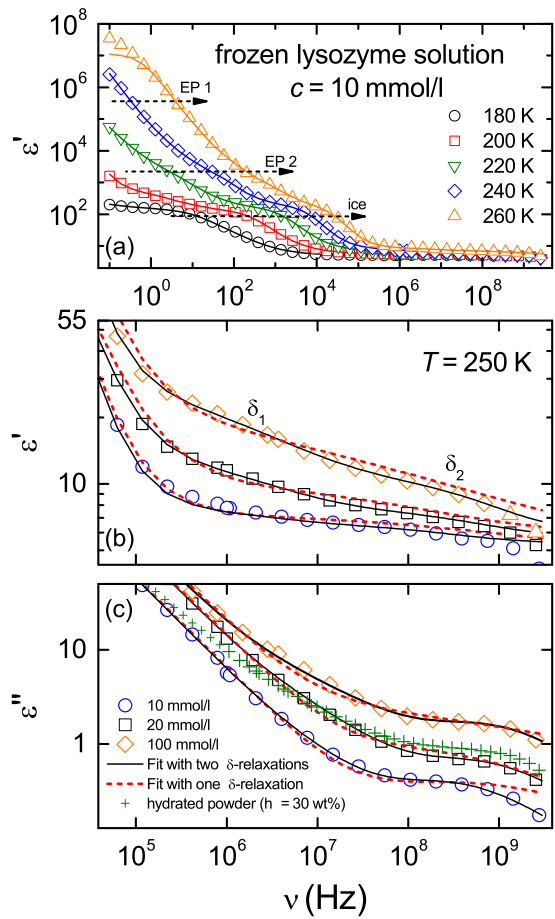}%
\caption{\label{fig2}Dielectric spectra of frozen protein solutions. (a) Dielectric constant of a 10\,mmol/l lysozyme solution measured at different temperatures below 273\,K. The solid lines are fits using the sum of five Cole-Cole functions. (b, c) Zoomed view of $\e'(\nu)$ (a) and $\e''(\nu)$ of differently concentrated protein solutions ($c=10$, 20, and 100\,mmol/l) measured at 250\,K. Lines are fits using five (solid lines) or four (dashed lines) Cole-Cole functions in total. Pluses in (c) represent the dielectric loss of a hydrated lysozyme powder ($h=30$\,wt\%).}
\end{figure}

Fig. \ref{fig2}(a) shows broadband spectra of the dielectric constant of a 10\,mmol/l lysozyme solution measured at different temperatures \textit{below} the freezing point of water [for $\varepsilon''(\nu)$ see \cite{supp}]. Starting at a low-frequency value of $10^7$, $\varepsilon'(\nu)$ at 260\,K drops to a value of the order of 10 at $\nu\approx5$\,MHz in three consecutive steps, indicating the existence of three relaxation processes (EP1, EP2, and ice).
Based on the unreasonably high dielectric strengths, relaxations EP1 ($\Delta\varepsilon>10^6$) and EP2 ($\Delta\varepsilon>10^3$) are attributed to EP effects often found in ionically conducting materials \cite{Macdonald1987a, Feldman2003, Pimenov1998, Gulich2009, Emmert2011, Wolf2011}.
In contrast, the third $\varepsilon'(\nu)$ step with $\Delta\varepsilon$ of the order of 100 is due to the typical relaxation process of ice arising from the mentioned proton-hopping processes \cite{Johari1981b,Gough1970,Shinyashiki2009}. With decreasing temperature, all these relaxation features shift to lower frequencies, directly mirroring the decreasing translational mobility of free ions and protons.
At frequencies higher than 500\,MHz, two additional relaxations are found. Hardly detectable in the scaling of Fig. \ref {fig2}(a), they become evident in the enlarged views of Figs. \ref{fig2}(b) and (c) showing $\varepsilon'$ and $\varepsilon''$ in the high-frequency range. These figures provide data of differently concentrated protein solutions measured at 250\,K. Obviously, there is a relaxation process just below 1\,GHz, revealed by a small step in $\varepsilon'$ (b) and a rather well-pronounced peak in $\varepsilon''$ (c). This relaxation, ``$\delta_2$'', is clearly seen for all shown concentrations. For the highest concentrated sample, 100\,mmol/l, faint indications for a second relaxation are found between $10^6$ and $10^7$ Hz. For the lower concentrated samples, this relaxation is less obvious but is clearly revealed by the fitting procedure described in the following.

The solid lines in Fig.~\ref{fig2} are fits using the sum of five Cole-Cole functions \cite{Cole1941}, $\varepsilon^{*}(\nu)=\sum_{n} \{\varepsilon_{\mathrm{\infty}}+{\Delta\varepsilon_n}/[{1+(\mathrm{i}\omega\tau_n)^{1-\alpha_n}]}\}$, to account for the five relaxations. Here $\tau$ and $\Delta\varepsilon$ represent the relaxation time and the dielectric strength, respectively. This function is an empirical extension of the Debye formula ($\alpha=0$) \cite{Debye1929}; the additional parameter $\alpha_n$ ($0\leq\alpha_n<1$) causes a symmetric broadening of the relaxation peaks.
For comparison, Figs. \ref{fig2}(b) and (c) also show fits with four Cole-Cole functions, i.e. using only one relaxation function to account for the $\delta$-dispersion range (dashed lines). In contrast to solutions above the freezing point \cite{Wolf2012}, where the $\beta$- and $\gamma$-relaxations partly superimpose the $\delta$-dispersion, obviously a satisfying description of the present frozen-solution data only is possible when assuming two $\delta$-relaxations. This is a clear hint at the bimodality of the hydration-shell dynamics. As expected for hydration-water related relaxations, the amplitudes of the $\delta$-relaxations increase with increasing protein content, i.e. with increasing number of bound water molecules. The pluses in Fig. \ref{fig2}(c) represent $\e''(\nu)$ of a hydrated lysozyme powder sample with a hydration degree of $h=30$\,wt\%. The dielectric loss of this sample obviously resembles that of the 20\,mmol/l solution.  While the protein content of this sample is significantly higher than for the 20\,mmol/l solution, the amount of water is clearly lower. Overall, this causes a $\delta_2$-relaxation of similar dielectric strength. The properties of the $\delta_1$-relaxation cannot be judged by eye, but are deduced and discussed below.

The most important parameter derived from the fits to the experimental data is the relaxation time $\tau$, characterizing the dynamics of the relaxing entities. In the case of thermally activated behavior, the temperature dependence of $\tau$ can be described by the Arrhenius law, $
\tau=\tau_{0}\exp[E_{\tau}/(k_{\mathrm{B}}T)]$, where $\tau_{0}$ is the inverse attempt frequency (typically of the order of phonon frequencies) and $E_\tau$ is the hindering barrier. In disordered matter, relaxation processes often show super-Arrhenius behavior, which can be described by the empirical Vogel-Fulcher-Tammann (VFT) formula, $\tau=\tau_{0}\exp[DT_\mathrm{VFT}/(T-T_{\mathrm{VFT}})]$ \cite{Angell1985, Angell1988}. Here $D$ is the so-called strength parameter. Large or small $D$ values imply small or strong deviations from thermally activated behavior, which is termed ``strong'' or ``fragile'' behavior, respectively \cite{Angell1985, Boehmer1993}.

\begin{figure}
\includegraphics[width=0.85\linewidth]{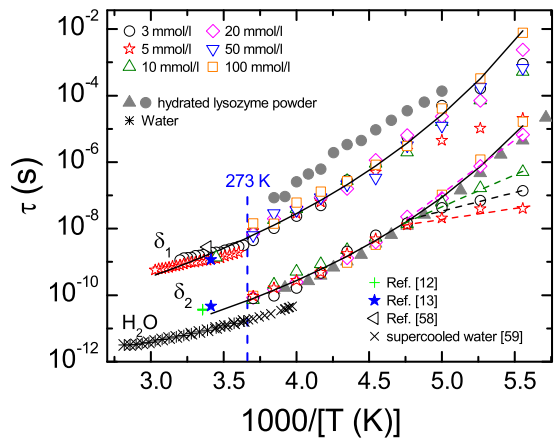}%
\caption{\label{fig3}Relaxation times of the $\delta_1$- and $\delta_2$-relaxations of differently concentrated protein solutions ($c=3$--100\,mmol/l) below the freezing point obtained in the present work. For comparison, literature data on $\tau_{\delta_1}$ \cite{Wolf2012,Cametti2011,Miura1994} and $\tau_{\delta_2}$ \cite{Oleinikova2004,Cametti2011} at $T>273$~K are included (Refs. \cite{Wolf2012,Cametti2011,Miura1994}: lysozyme solutions, Ref. \cite{Oleinikova2004}: ribonuclease A). The full circles and triangles are the relaxation times of the two intrinsic relaxations found in a hydrated lysozyme powder \cite{WolfDiss} ($h=30$\,wt\%). The asterisks and crosses are the relaxation times of water (own data) and supercooled water \cite{Bertolini1982}, respectively. The lines are master curves using the VFT formula.}
\end{figure}

The temperature dependence of the obtained relaxation times $\tau_{\delta_1}$ and $\tau_{\delta_2}$ of \textit{frozen} protein solutions ($c=3$--$100$\,mmol/l), is shown in Fig. \ref{fig3} (right of the vertical line marking the freezing point of water). In this Arrhenius presentation, a linear behavior reveals thermally activated behavior, whereas a curved temperature dependence is typical for fragile materials.
This figure also includes the relaxation times of the $\delta_1$-relaxation of 3 and 5\,mmol/l lysozyme solutions \textit{above} the freezing point, as published in \cite{Wolf2012} (left of dashed vertical line), literature values of the $\delta$-relaxations as reported in \cite{Miura1994, Oleinikova2004, Cametti2011} (different protein solutions), the relaxation times of pure water above the freezing point (own measurements) and of supercooled water from \cite{Bertolini1982}, as well as the relaxation times of the hydrated lysozyme powder with $h=30$\,wt\%. Obviously, the relaxation times $\tau_{\delta_1}$, derived from frozen protein solutions, nicely match those determined from solutions in the liquid state. The minor deviations (discontinuity around 273\,K) for the $\delta_1$-relaxation can be explained by the fact that the protein solutions above freezing point were fitted with only one $\delta$-relaxation as explained above \cite{Wolf2012}, whereas two relaxation functions were needed to describe the $\delta$-dispersion in case of the frozen protein solutions. This causes a slight spectral shift of the $\delta_1$-relaxation in the liquid solutions to higher frequencies, i.e. lower relaxation times. For the faster relaxation, $\delta_2$, the relaxation times of the frozen solutions are in very good accordance with the relaxation times reported for different protein solutions (ribonuclease A \cite{Oleinikova2004} and lysozyme \cite{Cametti2011}). The present results thus prove the proposed bimodality of the protein $\delta$-dispersion.

Some researches believe that both $\delta$-relaxations are strongly correlated with the protein hydration shell \cite{Cametti2011, Bone1992, Grant1986}. However, for the slower relaxation ($\delta_1$), protein-water collective motions or internal protein motions are alternative explanations \cite{Bone1982, Oleinikova2004, Grant1978}. Our own studies on hydrated proteins \cite{WolfDiss, Wolf2014tobe} show that this relaxation strongly depends on the degree of hydration. This is confirmed by the fact that the $\delta_1$-relaxation of the hydrated powder ($h=30\,$wt\%) (full circles in Fig. \ref{fig3}) is significantly slower than the $\delta_1$-relaxation of the fully hydrated protein solutions. As water is known to have a ``lubricating'' effect on proteins \cite{Pethig1994, Bone1982}, it seems reasonable to ascribe this hydration-dependent relaxation to a correlated protein-water movement. In contrast, the $\delta_2$-relaxation times of the hydrated protein powder (full triangles) agree with those of the frozen protein solutions. The dielectric strengths of the $\delta_2$-relaxation of the protein powder and of the 20\,mmol/l solutions are similar [cf. Fig. \ref{fig2}(c)]. This indicates that this relaxation not only depends on the amount of protein but also on the content of water in the sample. Moreover, it was found that this relaxation disappears when drying the protein powder \cite{WolfDiss}. These facts clearly suggest that the $\delta_2$-relaxation arises from loosely bound hydration water.

To emphasize the fragile temperature characteristics of the relaxation times, VFT-curves are drawn as solid lines in Fig. \ref{fig3}.
For the $\delta_1$-relaxation, the dynamics of nearly all samples follow this ``master curve'' ($\tau_0=3.6\times10^{-13}$\,s, $D=12.7$, $T_{\mathrm{VFT}}=117.5$\,K) throughout the whole temperature range. However, there are slight deviations from this behavior at the lowest temperatures investigated, especially for the 5\,mmol/l solution. These deviations show no systematic development with concentration and can be explained by the fact that the determination of $\tau_{\delta1}$ tends to become increasingly difficult with decreasing temperature, i.e. its uncertainty increases.

The relaxation times of the $\delta_2$-relaxation behave differently. At high temperatures, they follow a master curve ($\tau_0=2.4\times10^{-14}$\,s, $D=10.3$, $T_{\mathrm{VFT}}=119$\,K), but at a temperature of approximately 210\,K, the behavior of most samples changes to an Arrhenius temperature dependence, marked by the dashed lines. Such a crossover is often referred to as fragile-to-strong transition.
As the $\delta_2$-relaxation is ascribed to the dynamics of the loosely bound hydration water \cite{Cametti2011, Bone1982, Grant1986, Oleinikova2004}, this transition could be interpreted as the proposed fragile-to-strong transition of water in the No Man's Land \cite{Ito1999, Angell2008}, which is believed to accompany the structural change between the high-density and the low-density phase of water \cite{Chen2006}.
Such a dynamic crossover was also found, e.g., by Chen \textit{et al}. for fully hydrated lysozyme powder using neutron-scattering measurements \cite{Chen2006} and for confined water \cite{Ridi2009, Liu2006, Zanotti2005}. In the present work, this transition becomes less prominent with increasing protein concentration (Fig. \ref{fig3}), which might be explained by the fact that no complete hydration shell is formed for the samples with the highest protein concentrations (due to the lack of water) and thus the transition is suppressed.
However, it has to be stressed that the significance of the fragile-to-strong transition found in the present work is limited as $\tau_{\delta_2}$ has high uncertainty, especially at low temperatures and concentrations (cf. Fig. \ref{fig2}) (see \cite{supp} for a further check of the significance of the transition).

In summary, the present study demonstrates the great value of dielectric measurements on frozen protein solutions. With the help of these studies, the existence of a second $\delta$-relaxation could be unequivocally proven. Moreover, comparing the results with measurements on lysozyme solutions above the freezing point and with hydrated powders enables a clear assignment of the $\delta_2$-relaxation to loosely bound hydration water. For the $\delta_1$-relaxation, due to its strong dependence on hydration, a collective protein-water motion seems most probable.
In addition, indications for a fragile-to-strong transition were found. This dynamic change turned out to depend on the protein concentration of the solutions, becoming less prominent for high protein concentrations. While these findings are of limited significance only, further studies in frozen solutions of different proteins and with higher resolution seem a promising approach to help solving the long-standing problem of the fragile-to-strong transition.

\bibliography{FrozFinal}

\end{document}